\documentclass[12pt]{article}

\author{Yu.~M.~Zinoviev
       \thanks{E-mail address: Yurii.Zinoviev@ihep.ru} \\[0.5cm]
        {\it Institute for High Energy Physics} \\
        {\it of National Research Center "Kurchatov Institute"} \\
        {\it Protvino, Moscow Region, 142280, Russia}}
\title{Massless spin 2 interacting with massive higher spins in $d=3$}

\date{}

\begin{document}

\maketitle

\begin{abstract}
In this paper we consider massless spin 2 interacting with the massive
arbitrary spin fermions in $d=3$. First of all, we study all possible
deformations for the massive fermion unfolded equations in the
massless spin 2 background. We find three linearly independent
solutions one of which corresponds to the standard gravitational
interactions. Then for all three cases we reconstruct appropriate
Lagrangian formulation.
\end{abstract}

\thispagestyle{empty}
\newpage
\setcounter{page}{1}

\tableofcontents

\section{Introduction}

For a long time three dimensional space serves as a nice playground
for the investigation of higher spin interactions. The most well known
example is a Blencowe theory \cite{Ble89}, describing interactions for
infinite set of massless bosonic and fermionic fields. In the
metric-like formalism a complete classification of the cubic
interaction vertices for the bosonic fields was elaborated in
\cite{Mkr17,KM18}. For the spins $s \ge 2$ it was shown that for any
three spins $s_{1,2,3}$ satisfying a so-called triangular inequality 
$s_i + s_{i+1} > s_{i+2}$ there exist just one parity even and one
parity odd vertex. Using a frame-like formalism we confirmed these
results and extended them including massless fermions with spins 
$s \ge 3/2$ as well \cite{Zin21}. In the frame-like formalism it is
evident that it is not possible to construct any quartic or higher
order vertices. In spite of this fact, there exists a large number of
closed models with a finite number of higher spin fields (see e.g. 
\cite{CFPT10,CLW12,Tan12,Zin14a}). 

Recently, a complete classification for the cubic vertices for massive
fields was developed in the light-cone formalism \cite{Met20} (see
also
\cite{STT20}). The main result is that for any three spins there
exists at least one (sometimes two) cubic vertex. At the same time, in
our analyse based on the frame-like formalism \cite{Zin21} the
triangular inequality also appeared. The Lorentz covariant formulation
of the complete set of cubic vertices for the massive fields and their
relation with the classification of \cite{Met20} is still an open
question.

One more interesting and important case --- interactions of massless
and massive higher spins. Till now the most important result is a
Prokushkin-Vasiliev theory \cite{PV98,PSV99} which contain a complete
set of non-linear unfolded equations describing interaction of the
infinite set of massless bosonic and fermionic fields with massive
scalars and spinors. In \cite{Zin21} working both with Lagrangian and
unfolded formulations we consider possible generalization to the
massive higher spins. Once again we have found that the triangular
inequality is necessary.

Our analysis was based on the frame-like gauge invariant formalism for
the massive bosonic and fermionic fields \cite{BSZ12a,BSZ14a}, where
gauge invariance is achieved due to the introduction of appropriate
number of Stueckelberg fields. In such approach there appears a huge
ambiguity related with the (higher derivative) field redefinitions
\cite{BDGT18,KhZ21,KhZ21a} when one and the same theory may appear in
its non-abelian, abelian or trivially gauge invariant incarnations.
In-particular, we faced this problem trying to construct a Lagrangian
formulation for massive higher spin supermultiplet. The solution found
was based on the usage of the unfolded formulation for the massive
higher spins \cite{BPSS15,Zin15}. Recall, that such formulation
contains three sets of unfolded equations: for the one-forms, for the
Stueckelberg zero-forms and for the gauge invariant zero-forms. In
this
approach to find supertransformations one has to consider a
deformation of the unfolded equations in the background of massless
spin 3/2 field. In the sector of the gauge invariant zero-forms such
procedure appears to be non-ambiguous and consistency allows one to
promote these results on two other sectors \cite{BSZ16}. After that we
manged to reconstruct the Lagrangian formulation for these
supermultiplets \cite{BSZ17,BSZ17a}.

In this work we follow the same strategy for the case of massless spin
2 field interacting with the massive higher spins. Namely, we begin
with the deformation of the unfolded equations in the massless spin 2
background in the sector of the gauge invariant zero-forms. We found
three solutions one of which (naturally) corresponds to the standard
gravitational interaction. Then by consistency we promote these
solution to the two other sectors. At last we show that in all three
cases it is possible to reconstruct a Lagrangian formulation for the
appropriate cubic vertices. The gravitational interactions for massive
higher spin bosons were considered previously \cite{Zin12,BSZ12b}, so
in this work we restrict ourselves with the fermionic fields only.

The layout of the paper is simple. In Section 2 we provide all
necessary information on the Lagrangian and unfolded formulation for
the massive higher spin fermions. Section 3 devoted to the
deformation of the unfolded equations, while in Section 4 Lagrangian
formulation is reconstructed.\\
{\bf Notation and conventions} We work in the frame-like multispinor
formalism where all objects are forms having a number of completely
symmetric spinor indices (see \cite{BSZ17a} for details). A coordinate
free description of flat three dimensional space is given by the
background frame one-form $e^{\alpha(2)}$ and a background Lorentz
covariant derivative $D$ such that
$$
D \wedge D = 0, \qquad D \wedge e^{\alpha(2)} = 0.
$$
Also we use two and three forms defined as
$$
e^{\alpha(2)} \wedge e^{\beta(2)} = \varepsilon^{\alpha\beta}
E^{\alpha\beta}, \qquad E^{\alpha(2)} \wedge e^{\beta(2)} =
\varepsilon^{\alpha\beta} \varepsilon^{\alpha\beta} E.
$$
The massless spin 2 is described by a physical one-form
$H^{\alpha(2)}$
and an auxiliary one-form $\Omega^{\alpha(2)}$ with the free
Lagrangian
\begin{equation}
{\cal L}_0 = \Omega_{\alpha\beta} e^\beta{}_\gamma
\Omega^{\alpha\gamma} + \Omega_{\alpha(2)} D H^{\alpha(2)},
\end{equation}
which is invariant under the following gauge transformations
\begin{equation}
\delta \Omega^{\alpha(2)} = D \eta^{\alpha(2)}, \qquad
\delta H^{\alpha(2)} = D \xi^{\alpha(2)} + e^\alpha{}_\beta
\Omega^{\alpha\beta}.
\end{equation}
Equations of motion coincide with the gauge invariant curvatures
\begin{equation}
{\cal R}^{\alpha(2)} = D \Omega^{\alpha(2)}, \qquad
{\cal T}^{\alpha(2)} = D H^{\alpha(2)} + e^\alpha{}_\beta 
\Omega^{\alpha\beta}.
\end{equation}

\section{Massive  spin-(s+1/2) fermion}

Gauge invariant formulation of the massive spin-($s+1/2$) fermions in
$d=3$ \cite{BSZ14a,BSZ17} requires a set of one-forms 
$\Phi^{\alpha(2k+1)}$,  $0 \le k \le s-1$ and  zero-form 
$\phi^\alpha$. The free Lagrangian:
\begin{eqnarray}
\frac{1}{i} {\cal L}_0 &=& \sum_{k=0}^{s-1} \frac{(-1)^{k+1}}{2}
\Phi_{\alpha(2k+1)} D \Phi^{\alpha(2k+1)} + \frac{1}{2} \phi_\alpha
E^\alpha{}_\beta D \phi^\beta \nonumber \\
 && + \sum_{k=1}^{s-1} (-1)^{k+1} a_k \Phi_{\alpha(2k-1)\beta(2)}
e^{\beta(2)} \Phi^{\alpha(2k-1)} + a_0 \Phi_\alpha E^\alpha{}_\beta
\phi^\beta \nonumber \\
 && + \sum_{k=0}^{s-1} (-1)^{k+1} \frac{b_k}{2} \Phi_{\alpha(2k)\beta}
e^\beta{}_\gamma \Phi^{\alpha(2k)\gamma} - \frac{3b_0}{2} E
\phi_\alpha \phi^\alpha .
\end{eqnarray}
Here
\begin{equation}
b_k = \frac{(2s+1)}{(2k+3)}M, \qquad
a_k{}^2 = \frac{(s+k+1)(s-k)}{2(k+1)(2k+1)}M^2, \qquad
a_0{}^2 = 2s(s+1)M^2.
\end{equation}
This Lagrangian is invariant under the following gauge transformations
with the fermionic parameters $\zeta^{\alpha(2k+1)}$, 
$0 \le k \le s-1$:
\begin{eqnarray}
\delta_0 \Phi^{\alpha(2s-1)} &=& D \zeta^{\alpha(2s-1)} 
 + \frac{b_{s-1}}{(2s-1)} e^\alpha{}_\beta \zeta^{\alpha(2s-2)\beta}
 + \frac{a_{s-1}}{(2-1)(2s-1)} e^{\alpha(2)} \zeta^{\alpha(2s-3)},
\nonumber \\
\delta_0 \Phi^{\alpha(2k+1)} &=& D \zeta^{\alpha(2k+1)} +
\frac{b_k}{(2k+1)} e^\alpha{}_\beta \zeta^{\alpha(2k)\beta} \nonumber
\\
 && + \frac{a_k}{k(2k+1)} e^{\alpha(2)} \zeta^{\alpha(2k-1)}
+ a_{k+1} e_{\beta(2)} \zeta^{\alpha(2k+1)\beta(2)},  \\
\delta_0 \Phi^\alpha &=& D \zeta^\alpha + b_0 e^\alpha{}_\beta
\zeta^\beta + a_1 e_{\beta(2)} \zeta^{\alpha\beta(2)},  \nonumber \\
\delta_0 \phi^\alpha &=& a_0 \zeta^\alpha.   \nonumber
\end{eqnarray}
Note that such formalism admits an infinite spin limit $s \to \infty$,
$M \to 0$, $Ms = \mu = const$ \cite{Zin17} (see \cite{BS17} for
review):
\begin{equation}
b_k = \frac{2\mu}{(2k+3)}, \qquad
a_k{}^2 = \frac{\mu^2}{2(k+1)(2k+1)}, \qquad
a_0{}^2 = 2\mu^2.
\end{equation}
Note also that in some cases it appears to be convenient to work with
partially gauge fixed version then one sets $\phi^\alpha=0$. In this
case the Lagrangian is still invariant under the transformations with
the parameters $\zeta^{\alpha(2k+1)}$, $1 \le k \le s-1$, while
instead of $\zeta^\alpha$-invariance we have a constraint:
\begin{equation}
0 \approx D \frac{\delta{\cal L}_0}{\delta  \Phi_\alpha} + b_0 
e^\alpha{}_\beta \frac{\delta {\cal L}_0}{\delta \Phi_\beta} - a_1
e_{\beta(2)} \frac{\delta {\cal L}_0}{\delta  \Phi_{\alpha\beta(2)}}
 = - a_0{}^2 E^\alpha{}_\beta \Phi^\beta. 
\end{equation}
In \cite{KhZ22} we proved that such formalism does describe one
physical degree of freedom with mass $M$ and helicity $s+1/2$.

For each one-form $\Phi^{\alpha(2k+1)}$ there exists a gauge invariant
two-form ${\cal F}^{\alpha(2k+1)}$:
\begin{eqnarray}
{\cal F}^{\alpha(2k+1)} &=& D \Phi^{\alpha(2k+1)} + \frac{b_k}{(2k+1)}
e^\alpha{}_\beta \Phi^{\alpha(2k)\beta} \nonumber \\
 && + \frac{a_k}{k(2k+1)} e^{\alpha(2)} \Phi^{\alpha(2k-1)}
+ a_{k+1} e_{\beta(2)} \Phi^{\alpha(2k+1)\beta(2)}, \\
{\cal F}^\alpha &=& D \Phi^\alpha + b_0 e^\alpha{}_\beta
\Phi^\beta + a_1 e_{\beta(2)} \Phi^{\alpha\beta(2)} - a_0
E^\alpha{}_\beta \phi^\beta. \nonumber
\end{eqnarray}
But to construct a gauge invariant one-form for the Stueckelberg 
zero-form $\phi^\alpha$ one has introduce an extra zero-form
$\phi^{\alpha(3)}$ playing the role of Stueckelberg field for the
$\zeta^{\alpha(3)}$ transformation
$$
\delta \phi^{\alpha(3)} = a_0 \zeta^{\alpha(3)}.
$$
Then, to construct a gauge invariant one-form for $\phi^{\alpha(3)}$
one has introduce an extra zero-form $\phi^{\alpha(5)}$ and so on. The
procedure stops at $\phi^{\alpha(2s-1)}$ so that the complete set of
gauge invariant one-forms looks like:
\begin{eqnarray}
{\cal C}^\alpha &=& D \phi^\alpha - a_0 \Phi^\alpha + b_0 
e^\alpha{}_\beta \phi^\beta + c_1 e_{\beta(2)} \phi^{\alpha\beta(2)},
\nonumber \\
{\cal C}^{\alpha(2k+1)} &=& D \phi^{\alpha(2k+1)} - a_0
\Phi^{\alpha(2k+1)} + \frac{b_k}{(2k+1)} e^\alpha{}_\beta
\phi^{\alpha(2k)\beta} \nonumber \\
 && + \frac{a_k}{k(2k+1)} e^{\alpha(2)} \phi^{\alpha(2k-1)}
+ a_{k+1} e_{\beta(2)} \phi^{\alpha(2k+1)\beta(2)}, \\
{\cal C}^{\alpha(2s-1)} &=& D \phi^{\alpha(2s-1)} - a_0
\Phi^{\alpha(2s-1)} + \frac{b_{s-1}}{(2s-1)} e^\alpha{}_\beta
\phi^{\alpha(2s-2)\beta} \nonumber \\
 && + \frac{a_{s-1}}{(s-1)(2s-1)} e^{\alpha(2)} \phi^{\alpha(2s-3)},
\nonumber
\end{eqnarray}
where
\begin{equation}
\delta \phi^{\alpha(2k+1)} = a_0 \zeta^{\alpha(2k+1)}.
\end{equation}

Unfolded equations for the massive fermions \cite{BSZ16} (see
\cite{BPSS15,Zin15} for the bosonic fields) look as follows. On-shell
all gauge invariant two-forms ${\cal F} \approx 0$ and all gauge
invariant one-forms ${\cal C} \approx 0$ except the highest one:
\begin{equation}
0 \approx {\cal C}^{\alpha(2s-1)} + e_{\beta(2)} 
\phi^{\alpha(2s-1)\beta(2)}.
\end{equation}
Here the zero-form $\phi^{\alpha(2s+1)}$ is the first representative
of an infinite chain of the gauge invariant zero-forms 
$\phi^{\alpha(2k+1)}$, $k \ge s$, with the corresponding unfolded
equations
\begin{equation}
0 = D \phi^{\alpha(2k+1)} + e_{\beta(2)} \phi^{\alpha(2k+1)\beta(2)}
+ c_k e^\alpha{}_\beta \phi^{\alpha(2k)\beta} + d_k e^{\alpha(2)}
\phi^{\alpha(2k-1)}, 
\end{equation}
where
\begin{equation}
c_k = \frac{(2s+1)M}{(2k+1)(2k+3)}, \qquad
d_k = - \frac{(k-s)(k+s+1)}{2k(k+1)(2k+1)^2}M^2.
\end{equation}
Thus the complete set of the unfolded equations contains three
sectors: for the one-forms, for the Stueckelberg zero-forms and for
the gauge invariant zero-forms.

\section{Deformation of the unfolded equations}

Recall that the Prokushkin-Vasiliev theory \cite{PV98,PSV99} contains
a complete system of the non-linear unfolded equations describing an
interaction of massive scalars and spinors with the infinite set of
the massless higher spin fields. In this section we consider a
deformation of the unfolded equations for the massive higher spin
fermions in the presence of massless spin-2 field as a very first step
towards a possible generalization of this theory. We begin with the
sector of the gauge invariant zero-forms and then try to promote the
results into the sectors of the Stueckelberg zero-forms and the
one-forms.

\subsection{Sector of the gauge invariant zero-forms}

Let us consider two fermions: $\phi$ with mass $M_1$  and spin $s_1$
and $\psi$ with mass $M_2$ and spin $s_2$. The most general ansatz
for the deformed unfolded equations of the first fermion has the form
(see \cite{Zin21} for general discussion):
\begin{eqnarray}
0 &=& D \phi^{\alpha(2k+1)} + e_{\beta(2)} \phi^{\alpha(2k+1)\beta(2)}
+ c_k e^\alpha{}_\beta \phi^{\alpha(2k)\beta} + d_k e^{\alpha(2)}
\phi^{\alpha(2k-1)} \nonumber \\
 && + f_{1,k} H_{\beta(2)} \psi^{\alpha(2k+1)\beta(2)}
+ f_{2,k} H^\alpha{}_\beta \psi^{\alpha(2k)\beta}
+ f_{3,k} H^{\alpha(2)} \psi^{\alpha(2k-1)} \nonumber \\
 && + g_{1,k} \Omega_{\beta(2)} \psi^{\alpha(2k+1)\beta(2)}
+ g_{2,k} \Omega^\alpha{}_\beta \psi^{\alpha(2k)\beta}
+ g_{3,k} \Omega^{\alpha(2)} \psi^{\alpha(2k-1)}.
\end{eqnarray}
Similarly for the second fermion with $\tilde{c}$, $\tilde{d}$,
$\tilde{g}$ and $\tilde{f}$. First of all, consistency requires that
masses must be equal $M_1=M_2=M$ while spins may be equal $s_1 = s_2$
or differ by one unit $s_1 = s_2 \pm 1$. We have found three linearly
independent solutions. \\
{\bf Case I --- gravity} The two spins are equal so we can restrict
ourselves with just one fermion. The solution can be written as
\begin{equation}
0 = \hat{D} \phi^{\alpha(2k+1)} + \hat{e}_{\beta(2)} 
\phi^{\alpha(2k+1)\beta(2)} + c_k \hat{e}^\alpha{}_\beta 
\phi^{\alpha(2k)\beta} + d_k \hat{e}^{\alpha(2)}
\phi^{\alpha(2k-1)},
\end{equation}
where
\begin{equation}
\hat{D} \phi^{\alpha(2k+1)} = D \phi^{\alpha(2k+1)} + g_0
\Omega^\alpha{}_\beta \phi^{\alpha(2k)\beta}, \qquad
\hat{e}^{\alpha(2)} = e^{\alpha(2)} + g_0 H^{\alpha(2)}. \label{def_1}
\end{equation}
This solution corresponds to the standard minimal gravitational
interaction (hence the name). Indeed, in the next approximation we
need
\begin{eqnarray}
0 &=& D \Omega^{\alpha(2)} + \frac{g_0}{2} \Omega^\alpha{}_\beta
\Omega^{\alpha\beta}, \nonumber \\
0 &=& D H^{\alpha(2)} + (e^\alpha{}_\beta + g_0 H^\alpha{}_\beta)
\Omega^{\alpha\beta}. 
\end{eqnarray}
{\bf Case II --- non-gravity 1} Here two spins are also equal and we
consider one fermion only. Then the solution has the form:
\begin{equation}
0 = D \phi^{\alpha(2k+1)} + \hat{e}_{\beta(2)} 
\phi^{\alpha(2k+1)\beta(2)} + c_k \hat{e}^\alpha{}_\beta 
\phi^{\alpha(2k)\beta} + d_k \hat{e}^{\alpha(2)}
\phi^{\alpha(2k-1)}, 
\end{equation}
where
\begin{equation}\label{def_2}
\hat{e}^{\alpha(2)} = e^{\alpha(2)} + g_0 \Omega^{\alpha(2)}. 
\end{equation}
No deformations for spin 2 equations in  this case. \\
{\bf Case III --- non-gravity 2} Now spins $s_1=s+1/2$ and
$s_2=s-1/2$. The solution:
\begin{eqnarray}
0 &=& D \phi^{\alpha(2k+1)} + e_{\beta(2)} \phi^{\alpha(2k+1)\beta(2)}
+ c_k e^\alpha{}_\beta \phi^{\alpha(2k)\beta} + d_k e^{\alpha(2)}
\phi^{\alpha(2k-1)} \nonumber \\
 && + g_0 \Omega_{\beta(2)} \psi^{\alpha(2k+1)\beta(2)}
+ g_{2,k} \Omega^\alpha{}_\beta \psi^{\alpha(2k)\beta}
+ g_{3,k} \Omega^{\alpha(2)} \psi^{\alpha(2k-1)}, \\
0 &=& D \psi^{\alpha(2k+1)} + e_{\beta(2)} \psi^{\alpha(2k+1)\beta(2)}
+ \tilde{c}_k e^\alpha{}_\beta \psi^{\alpha(2k)\beta} +
\tilde{d}_k e^{\alpha(2)} \psi^{\alpha(2k-1)} \nonumber \\
 && + \tilde{g}_0 \Omega_{\beta(2)} \phi^{\alpha(2k+1)\beta(2)} 
+ \tilde{g}_{2,k} \Omega^\alpha{}_\beta \phi^{\alpha(2k)\beta}
+ \tilde{g}_{3,k} \Omega^{\alpha(2)} \phi^{\alpha(2k-1)},
\end{eqnarray}
where
\begin{equation}
g_{2,k} = \frac{2(k+s+1)M}{(2k+1)(2k+3)}g_0, \qquad
g_{3,k} = \frac{(k+s)(k+s+1)M^2}{2k(k+1)(2k+1)^2}g_0,
\end{equation}
\begin{equation}
\tilde{g}_{2,k} = - \frac{2(k-s+1)M}{(2k+1)(2k+3)}\tilde{g}_0, \qquad
\tilde{g}_{3,k} = \frac{(k-s)(k-s+1)M^2}{2k(k+1)(2k+1)^2}\tilde{g}_0.
\end{equation}
The structure of quadratic terms corresponds to non-gravity 1, i.e we
need something like
\begin{equation}
0 = D \omega^{\alpha(2)} + \frac{g_0}{2} \Omega^\alpha{}_\beta
\Omega^{\alpha\beta}, \qquad 0 = D \Omega^{\alpha(2)}.
\end{equation}
Noe we consider these three cases in turn and promote the solutions to
the other sectors.

\subsection{Gravity}

In this case the deformed equations can be straightforwardly written
with the standard minimal substitution rules. For the one-forms we
obtain
\begin{eqnarray}
0 &=& \hat{D} \Phi^{\alpha(2k+1)} + \frac{b_k}{(2k+1)}
\hat{e}^\alpha{}_\beta \Phi^{\alpha(2k)\beta} \nonumber \\
 && + \frac{a_k}{k(2k+1)} \hat{e}^{\alpha(2)} \Phi^{\alpha(2k-1)}
+ a_{k+1} \hat{e}_{\beta(2)} \Phi^{\alpha(2k+1)\beta(2)}, \\
0 &=& \hat{D} \Phi^\alpha + b_0 \hat{e}^\alpha{}_\beta
\Phi^\beta + a_1 \hat{e}_{\beta(2)} \Phi^{\alpha\beta(2)} 
- \frac{a_0g_0}{4} (e^\alpha{}_\beta H^\beta{}_\gamma +
H^\alpha{}_\beta e^\beta{}_\gamma) \phi^\gamma, \nonumber
\end{eqnarray}
while the deformed equations for the Stueckelberg zero-forms look
like:
\begin{eqnarray}
0 &=& \hat{D} \phi^\alpha - a_0 \Phi^\alpha + b_0 
\hat{e}^\alpha{}_\beta \phi^\beta + a_1 \hat{e}_{\beta(2)}
\phi^{\alpha\beta(2)}, \nonumber \\
0 &=& \hat{D} \phi^{\alpha(2k+1)} - a_0
\Phi^{\alpha(2k+1)} + \frac{b_k}{(2k+1)} \hat{e}^\alpha{}_\beta
\phi^{\alpha(2k)\beta} \nonumber \\
 && + \frac{a_k}{k(2k+1)} \hat{e}^{\alpha(2)} \phi^{\alpha(2k-1)}
+ a_{k+1} \hat{e}_{\beta(2)} \phi^{\alpha(2k+1)\beta(2)}, \\
0 &=& \hat{D} \phi^{\alpha(2s-1)} - a_0
\Phi^{\alpha(2s-1)} + \frac{b_{s-1}}{(2s-1)} \hat{e}^\alpha{}_\beta
\phi^{\alpha(2s-2)\beta} \nonumber \\
 && + \frac{a_{s-1}}{(s-1)(2s-1)} \hat{e}^{\alpha(2)}
\phi^{\alpha(2s-3)} + \hat{e}_{\beta(2)} \phi^{\alpha(2s-1)\beta(2)}.
\nonumber
\end{eqnarray}
Here $\hat{D}$ and $\hat{e}$ are the same as in (\ref{def_1}).

\subsection{Non-gravity 1}

Similarly, for the deformed equations of the one-forms we obtain
\begin{eqnarray}
0 &=& D \Phi^{\alpha(2k+1)} + \frac{b_k}{(2k+1)}
\hat{e}^\alpha{}_\beta \Phi^{\alpha(2k)\beta} \nonumber \\
 && + \frac{a_k}{k(2k+1)} \hat{e}^{\alpha(2)} \Phi^{\alpha(2k-1)}
+ a_{k+1} \hat{e}_{\beta(2)} \Phi^{\alpha(2k+1)\beta(2)}, \\
0 &=& D \Phi^\alpha + b_0 \hat{e}^\alpha{}_\beta
\Phi^\beta + a_1 \hat{e}_{\beta(2)} \Phi^{\alpha\beta(2)} 
- \frac{a_0g_0}{4} (e^\alpha{}_\beta \Omega^\beta{}_\gamma +
\Omega^\alpha{}_\beta e^\beta{}_\gamma) \phi^\gamma, \nonumber
\end{eqnarray}
while the deformed equations for the Stueckelberg zero-forms look
like:
\begin{eqnarray}
0 &=& D \phi^\alpha - a_0 \Phi^\alpha + b_0 
\hat{e}^\alpha{}_\beta \phi^\beta + c_1 \hat{e}_{\beta(2)}
\phi^{\alpha\beta(2)}, \nonumber \\
0 &=& D \phi^{\alpha(2k+1)} - a_0
\Phi^{\alpha(2k+1)} + \frac{b_k}{(2k+1)} \hat{e}^\alpha{}_\beta
\phi^{\alpha(2k)\beta} \nonumber \\
 && + \frac{a_k}{k(2k+1)} \hat{e}^{\alpha(2)} \phi^{\alpha(2k-1)}
+ a_{k+1} \hat{e}_{\beta(2)} \phi^{\alpha(2k+1)\beta(2)}, \\
0 &=& D \phi^{\alpha(2s-1)} - a_0
\Phi^{\alpha(2s-1)} + \frac{b_{s-1}}{(2s-1)} \hat{e}^\alpha{}_\beta
\phi^{\alpha(2s-2)\beta} \nonumber \\
 && + \frac{a_{s-1}}{(s-1)(2s-1)} \hat{e}^{\alpha(2)}
\phi^{\alpha(2s-3)} + \hat{e}_{\beta(2)} \phi^{\alpha(2s-1)\beta(2)}.
\nonumber
\end{eqnarray}
Here $\hat{e}$ is the same as in (\ref{def_2}).

\subsection{Non-gravity 2}

In this case we have to consider two fields with different spins. Let
us begin with the following ansatz for the Stueckelberg zero-forms of
the first field:
\begin{eqnarray}
0 &=& D \phi^\alpha - a_0 \Phi^\alpha + b_0  e^\alpha{}_\beta
\phi^\beta + a_1 e_{\beta(2)} \phi^{\alpha\beta(2)} \nonumber \\
 && + h_{1,0} \Omega^\alpha{}_\beta \psi^\beta
 + h_{3,0} \Omega_{\beta(2)} \psi^{\alpha\beta(2)}, \nonumber \\
0 &=& D \phi^{\alpha(2k+1)} - a_0 \Phi^{\alpha(2k+1)} 
+ \frac{b_k}{(2k+1)} e^\alpha{}_\beta \phi^{\alpha(2k)\beta} 
\nonumber \\
 && + \frac{a_k}{k(2k+1)} e^{\alpha(2)} \phi^{\alpha(2k-1)}
+ a_{k+1} e_{\beta(2)} \phi^{\alpha(2k+1)\beta(2)} \nonumber \\
 && + h_{1,k} \Omega^\alpha{}_\beta \psi^{\alpha(2k)\beta}
+ h_{2,k} \Omega^{\alpha(2)} \psi^{\alpha(2k-1)} 
 + h_{3,k} \Omega_{\beta(2)} \psi^{\alpha(2k+1)\beta(2)}, \\
0 &=& D \phi^{\alpha(2s-1)} - a_0 \Phi^{\alpha(2s-1)} 
+ \frac{b_{s-1}}{(2s-1)} e^\alpha{}_\beta \phi^{\alpha(2s-2)\beta}
\nonumber \\
 && + \frac{a_{s-1}}{(s-1)(2s-1)} e^{\alpha(2)} \phi^{\alpha(2s-3)}
 + e_{\beta(2)} \phi^{\alpha(2s-1)\beta(2)} \nonumber \\
 && + h_{1,s-1} \Omega^\alpha{}_\beta \psi^{\alpha(2s-2)\beta}
+ h_{2,s-1} \Omega^{\alpha(2)} \psi^{\alpha(2s-3)}
+ h_{3,s-1} \Omega_{\beta(2)} \psi^{\alpha(2s-1)\beta(2)}. \nonumber
\end{eqnarray}
Note that the terms with the coefficients $h_{3,s-2}$, $h_{1,s-1}$ and
$h_{3,s-1}$ contain gauge invariant zero-forms of the second fields.
Consistency requirement gives:
\begin{eqnarray}
h_{1,k} &=& \frac{2\sqrt{2(s+k+1)(s-k-1)}}{(2k+1)(2k+3)}M^2g_0, \qquad
h_{1,s-1} = \frac{4Msg_0}{(2s-1)(2s+1)}, \nonumber \\
h_{2,k} &=& \frac{1}{k(2k+1)}
\sqrt{\frac{(s+k+1)(s+k)}{(k+1)(2k+1)}}M^2g_0, \\
h_{3,k} &=& \sqrt{\frac{(s-k-2)(s-k-1)}{(k+2)(2k+3)}}M^2g_0, \qquad
h_{3,s-2} = a_{s-1}g_0, \qquad
h_{3,s-1} = g_0. \nonumber
\end{eqnarray}
This in turn requires the following deformations for the one-forms:
\begin{eqnarray}
0 &=& D \Phi^{\alpha(2s-1)} + \frac{b_{s-1}}{(2s-1)}
e^\alpha{}_\beta \Phi^{\alpha(2s-2)\beta} 
+ \frac{a_{s-1}}{(s-1)(2s-1)} e^{\alpha(2)} \Phi^{\alpha(2s-3)},
\nonumber \\
 && + \frac{\tilde{a}_0}{a_0} h_{2,s-2} \Omega^{\alpha(2)}
\Psi^{\alpha(2s-3)}, \nonumber \\
0 &=& D \Phi^{\alpha(2k+1)} + \frac{b_k}{(2k+1)}
e^\alpha{}_\beta \Phi^{\alpha(2k)\beta} \nonumber \\
 && + \frac{a_k}{k(2k+1)} e^{\alpha(2)} \Phi^{\alpha(2k-1)}
+ a_{k+1} e_{\beta(2)} \Phi^{\alpha(2k+1)\beta(2)} \\
 && + \frac{\tilde{a}_0}{a_0} [ h_{1,k} \Omega^\alpha{}_\beta
\Psi^{\alpha(2k)\beta} + h_{2,k} \Omega^{\alpha(2)}
\Psi^{\alpha(2k-1)} + h_{3,k} \Omega_{\beta(2)}
\Psi^{\alpha(2k+1)\beta(2)} ], \nonumber \\
0 &=& D \Phi^\alpha + b_0 e^\alpha{}_\beta
\Phi^\beta + a_1 e_{\beta(2)} \Phi^{\alpha\beta(2)} - a_0
E^\alpha{}_\beta \phi^\beta \nonumber \\
 && + \frac{\tilde{a}_0}{a_0} [ h_{1,0} \Omega^\alpha{}_\beta
\Psi^\beta + h_{3,0} \Omega_{\beta(2)} \Psi^{\alpha\beta(2)}]
- \frac{\tilde{a}_0Mg_0}{4} (e^\alpha{}_\beta \Omega^\beta{}_\gamma
+ \Omega^\alpha{}_\beta e^\beta{}_\gamma) \psi^\gamma. \nonumber
\end{eqnarray}

Similarly, the ansatz for the Stueckelberg zero-forms of the second
field:
\begin{eqnarray}
0 &=& D \psi^\alpha - \tilde{a}_0 \Psi^\alpha + \tilde{b}_0 
e^\alpha{}_\beta \psi^\beta + \tilde{a}_1 e_{\beta(2)}
\psi^{\alpha\beta(2)} \nonumber \\
 && + \tilde{h}_{1,0} \Omega^\alpha{}_\beta \phi^\beta
 + \tilde{h}_{3,0} \Omega_{\beta(2)} \phi^{\alpha\beta(2)}, 
\nonumber \\
0 &=& D \psi^{\alpha(2k+1)} - \tilde{a}_0 \Psi^{\alpha(2k+1)} 
+ \frac{\tilde{b}_k}{(2k+1)} e^\alpha{}_\beta \psi^{\alpha(2k)\beta} 
\nonumber \\
 && + \frac{\tilde{a}_k}{k(2k+1)} e^{\alpha(2)} \psi^{\alpha(2k-1)}
+ \tilde{a}_{k+1} e_{\beta(2)} \psi^{\alpha(2k+1)\beta(2)} 
\nonumber \\
 && + \tilde{h}_{1,k} \Omega^\alpha{}_\beta \phi^{\alpha(2k)\beta}
+ \tilde{h}_{2,k} \Omega^{\alpha(2)} \phi^{\alpha(2k-1)} 
 + \tilde{h}_{3,k} \Omega_{\beta(2)} \phi^{\alpha(2k+1)\beta(2)}, \\
0 &=& D \psi^{\alpha(2s-3)} - \tilde{a}_0 \Psi^{\alpha(2s-3)} 
+ \frac{\tilde{b}_{s-2}}{(2s-3)} e^\alpha{}_\beta 
\psi^{\alpha(2s-4)\beta} \nonumber \\
 && + \frac{\tilde{a}_{s-2}}{(s-2)(2s-3)} e^{\alpha(2)}
\psi^{\alpha(2s-5)} + e_{\beta(2)} \psi^{\alpha(2s-3)\beta(2)}
\nonumber \\
 && + \tilde{h}_{1,s-2} \Omega^\alpha{}_\beta \phi^{\alpha(2s-4)\beta}
+ \tilde{h}_{2,s-2} \Omega^{\alpha(2)} \phi^{\alpha(2s-5)}
+ \tilde{h}_{3,s-2} \Omega_{\beta(2)} \phi^{\alpha(2s-3)\beta(2)}.
\nonumber
\end{eqnarray}
Note that all terms with the coefficients $\tilde{h}$ contain
Stueckelberg zero forms of the first field. We obtain:
\begin{eqnarray}
\tilde{h}_{1,k} &=& \frac{2\sqrt{(s+k+1)(s-k-1)}}{(2k+1)(2k+3)}
\tilde{g}_0, \nonumber \\
\tilde{h}_{2,k} &=& \frac{1}{k(2k+1)} 
\sqrt{\frac{(s-k-1)(s-k)}{2(k+1)(2k+1)}} \tilde{g}_0, \qquad k > 0, \\
\tilde{h}_{3,k} &=& \sqrt{\frac{(s+k+2)(s+k+1)}{2(k+2)(2k+3)}}
\tilde{g}_0. \nonumber 
\end{eqnarray}
This in turn requires the following deformations for the one-forms:
\begin{eqnarray}
0 &=& D \Psi^{\alpha(2k+1)} + \frac{\tilde{b}_k}{(2k+1)}
e^\alpha{}_\beta \Psi^{\alpha(2k)\beta} \nonumber \\
 && + \frac{\tilde{a}_k}{k(2k+1)} e^{\alpha(2)} \Psi^{\alpha(2k-1)}
+ \tilde{a}_{k+1} e_{\beta(2)} \Psi^{\alpha(2k+1)\beta(2)} 
\nonumber \\
 && + \frac{a_0}{\tilde{a}_0} [ \tilde{h}_{1,k} \Omega^\alpha{}_\beta
\Phi^{\alpha(2k)\beta} + \tilde{h}_{2,k} \Omega^{\alpha(2)}
\Phi^{\alpha(2k-1)} + \tilde{h}_{3,k} \Omega_{\beta(2)}
\Phi^{\alpha(2k+1)\beta(2)} ], \\
0 &=& D \Psi^\alpha + \tilde{b}_0 e^\alpha{}_\beta
\Psi^\beta + \tilde{a}_1 e_{\beta(2)} \Psi^{\alpha\beta(2)} - 
\tilde{a}_0 E^\alpha{}_\beta \psi^\beta \nonumber \\
 && + \frac{a_0}{\tilde{a}_0} [ \tilde{h}_{1,0} \Omega^\alpha{}_\beta
\Phi^\beta + \tilde{h}_{3,0} \Omega_{\beta(2)} \Phi^{\alpha\beta(2)}]
- \frac{a_0\tilde{g}_0}{4M} (e^\alpha{}_\beta \Omega^\beta{}_\gamma
+ \Omega^\alpha{}_\beta e^\beta{}_\gamma) \phi^\gamma. \nonumber
\end{eqnarray}

\section{Cubic vertices}

In this section we reconstruct the Lagrangian formulations for all
three cases above. Recall that the free Lagrangian contains the 
one-forms $\Phi^{\alpha(2k+1)}$, $0 \le k \le s-1$ and one zero-form
$\phi^\alpha$ only, while all other zero-forms are interpreted as the
higher derivatives of the Lagrangian fields. Analyzing the unfolded
equations obtained above, we see  that for the gravity case it is
enough to consider terms with no more than one derivative (taking into
account that the auxiliary field $\Omega^{\alpha(2)}$ is equivalent
to the first derivative of the physical field $H^{\alpha(2)}$), while
in the two non-gravity cases we have to consider terms with up to two
derivatives.

\subsection{Gravity}

In this case the cubic vertex (we set $g_0=1$ for simplicity)  can be
written as
\begin{eqnarray}
\frac{1}{i} {\cal L}_1 &=& \sum_{k=0}^{s-1} (-1)^{k+1} 
[\frac{(2k+1)}{2} \Phi_{\alpha(2k)\beta} \Omega^\beta{}_\gamma
\Phi^{\alpha(2k)\gamma} + \frac{b_k}{2} \Phi_{\alpha(2k)\beta}
H^\beta{}_\gamma \Phi^{\alpha(2k)\gamma}] \nonumber \\
 && + \sum_{k=1}^{s-1} (-1)^{k+1} a_k \Phi_{\alpha(2k-1)\beta(2)}
H^{\beta(2)} \Phi^{\alpha(2k-1)} \nonumber \\
 && - \frac{1}{4} (E\Omega) \phi_\alpha
\phi^\alpha + \frac{1}{8} \phi_\alpha [ e^\alpha{}_\beta 
H^\beta{}_\gamma + H^\alpha{}_\beta e^\beta{}_\gamma] D \phi^\gamma
\nonumber \\
 && + \frac{a_0}{4} \Phi_\alpha [ e^\alpha{}_\beta H^\beta{}_\gamma +
H^\alpha{}_\beta e^\beta{}_\gamma] \phi^\gamma - \frac{3b_0}{4} 
(EH) \phi_\alpha \phi^\alpha. 
\end{eqnarray}
All the gauge transformations (for the massless spin two as well as
for
massive fermion) requires non-trivial corrections.\\
{\bf $\eta^{\alpha(2)}$-transformations}  The gauge invariance
requires:
\begin{equation}
\delta \Phi^{\alpha(2k+1)} = - \eta^\alpha{}_\beta 
\Phi^{\alpha(2k)\beta}, \qquad \delta \phi^\alpha = -
\eta^\alpha{}_\beta \phi^\beta.
\end{equation}
{\bf $\xi^{\alpha(2)}$-transformations} Here we obtain:
\begin{eqnarray}
\delta \Phi^{\alpha(2k+1)} &=& - \frac{b_k}{(2k+1)} \xi^\alpha{}_\beta
\Phi^{\alpha(2k)\beta} - \frac{a_k}{k(2k+1)} \xi^{\alpha(2)}
\Phi^{\alpha(2k-1)} - a_{k+1} \xi_{\beta(2)} 
\Phi^{\alpha(2k+1)\beta(2)}, \nonumber \\
\delta \Phi^\alpha &=& - b_0 \xi^\alpha{}_\beta \Phi^\beta - a_1
\xi_{\beta(2)} \Phi^{\alpha\beta(2)} - \frac{a_0}{4}
(e^\alpha{}_\beta \xi^\beta{}_\gamma - \xi^\alpha{}_\beta
e^\beta{}_\gamma) \phi^\gamma, \\
\delta \phi^\alpha &=& - b_0 \xi^\alpha{}_\beta \phi^\beta
- a_1 \xi_{\beta(2)} \phi^{\alpha\beta(2)}. \nonumber 
\end{eqnarray}
{\bf $\zeta^{\alpha(2k+1)}$-transformations} The corrections for the
fermion look like:
\begin{eqnarray}
\delta \Phi^{\alpha(2k+1)} &=& \frac{b_k}{(2k+1)} H^\alpha{}_\beta
\zeta^{\alpha(2k)\beta} + \frac{a_k}{k(2k+1)} H^{\alpha(2)}
\zeta^{\alpha(2k-1)} + a_{k+1} H_{\beta(2)} 
\zeta^{\alpha(2k+1)\beta(2)}, \nonumber \\
\delta \Phi^\alpha &=& \Omega^\alpha{}_\beta \zeta^\beta + b_0
H^\alpha{}_\beta \zeta^\beta + a_1 H_{\beta(2)}
\zeta^{\alpha\beta(2)},
\end{eqnarray}
while for the massless spin-2 they have the form:
\begin{eqnarray}
\delta \Omega^{\alpha(2)} &=& \sum_{k=1}^{s-1} (-1)^{k+1} 
[ b_k \Phi^{\alpha\beta(2k)} \zeta^\alpha{}_{\beta(2k)}
+ a_k \Phi^{\alpha(2)\beta(2k-1)} \zeta_{\beta(2k-1)}
- a_k \Phi_{\beta(2k-1)} \zeta^{\alpha(2)\beta(2k-1)}] \nonumber  \\
 && - b_0 \Phi^\alpha \zeta^\alpha + \frac{a_0}{8}
(\phi^\alpha e^{\alpha\beta} \zeta_\beta + \phi_\beta
e^{\beta\alpha} \zeta^\alpha),\\
\delta H^{\alpha(2)} &=& \sum_{k=0}^{s-1} (-1)^{k+1} 
(2k+1) \Phi^{\alpha\beta(2k)} \zeta^\alpha{}_{\beta(2k)}. \nonumber
\end{eqnarray}
Note that the last formulas are completely consistent with corrections
to the massless spin-2 curvatures:
\begin{eqnarray}
\Delta {\cal R}^{\alpha(2)} &=& \sum_{k=0}^{s-1} (-1)^{k+1}
\frac{b_k}{2} \Phi^{\alpha\beta(2k)} \Phi^\alpha{}_{\beta(2k)}
+ \sum_{k=1}^{s-1} (-1)^{k+1} a_k \Phi^{\alpha(2)\beta(2k-1)} 
\Phi_{\beta(2k-1)} \nonumber \\
 && + \frac{1}{16} (\phi_\beta e^{\beta\alpha} D \phi^\alpha -
\phi^\alpha e^\alpha{}_\beta D \phi^\beta) - \frac{a_0}{8}
(\Phi_\beta e^{\beta\alpha} \psi^\alpha - \Phi^\alpha
e^\alpha{}_\beta \phi^\beta) - \frac{3b_0}{4} E^{\alpha(2)}
\phi_\beta \phi^\beta, \\
\Delta {\cal T}^{\alpha(2)} &=& \sum_{k=0}^{s-1} (-1)^{k+1}
\frac{(2k+1)}{2} \Phi^{\alpha\beta(2k)} \Phi^\alpha{}_{\beta(2k)} 
- \frac{1}{4} E^{\alpha(2)} \phi_\beta \phi^\beta. \nonumber
\end{eqnarray}
Note also that on partial gauge fixing $\phi^\alpha = 0$ the
invariance under the $\eta^{\alpha(2)}$ and 
$\zeta^{\alpha(2k+1)}$, $k>0$ transformations still remain,
$\xi^{\alpha(2)}$ transformations produce
$$
\delta {\cal L} = - \frac{a_0{}^2}{4} \Phi_\alpha e^\alpha{}_\beta
\xi^{\beta\gamma} \Phi_\gamma,
$$
while instead of $\zeta^\alpha$-invariance we obtain a constraint
\begin{equation}
0  = E^\alpha{}_\beta \Phi^\beta + \frac{1}{4} (e^\alpha{}_\beta
H^\beta{}_\gamma + H^\alpha{}_\beta e^\beta{}_\gamma) \Phi^\gamma.
\end{equation}

\subsection{Non-gravity 1}

In this case the cubic vertex (we also set $g_0=1$) can be written as
follows:
\begin{eqnarray}
\frac{1}{i} {\cal L}_1 &=& \sum_{k=0}^{s-1} (-1)^{k+1} 
\frac{b_k}{2} \Phi_{\alpha(2k)\beta} \Omega^\beta{}_\gamma
\Phi^{\alpha(2k)\gamma} + \sum_{k=1}^{s-1} (-1)^{k+1} a_k
\Phi_{\alpha(2k-1)\beta(2)} \Omega^{\beta(2)} \Phi^{\alpha(2k-1)}
\nonumber \\
 && + \frac{1}{8} \phi_\alpha 
[ e^\alpha{}_\beta \Omega^\beta{}_\gamma + \Omega^\alpha{}_\beta
e^\beta{}_\gamma] D \phi^\gamma 
 + \frac{a_0}{4} \Phi_\alpha [ e^\alpha{}_\beta
\Omega^\beta{}_\gamma + \Omega^\alpha{}_\beta e^\beta{}_\gamma]
\phi^\gamma - \frac{3b_0}{4} (E\Omega) \phi_\alpha \phi^\alpha.
\end{eqnarray}
The field $H^{\alpha(2)}$ does not enter the Lagrangian so the vertex
is trivially invariant under the $\xi^{\alpha(2)}$-transformations.\\
{\bf $\eta^{\alpha(2)}$-transformations} Here we need the following
corrections:
\begin{eqnarray}
\delta \Phi^{\alpha(2k+1)} &=& - \frac{b_k}{(2k+1)} 
\eta^\alpha{}_\beta \Phi^{\alpha(2k)\beta} - \frac{a_k}{k(2k+1)}
\eta^{\alpha(2)} \Phi^{\alpha(2k-1)} - a_{k+1} \eta_{\beta(2)} 
\Phi^{\alpha(2k+1)\beta(2)}, \nonumber \\
\delta \Phi^\alpha &=& - b_0 \eta^\alpha{}_\beta \Phi^\beta - a_1
\eta_{\beta(2)} \Phi^{\alpha\beta(2)} - \frac{a_0}{4}
(e^\alpha{}_\beta \eta^\beta{}_\gamma - \eta^\alpha{}_\beta
e^\beta{}_\gamma) \phi^\gamma, \\
\delta \phi^\alpha &=& - b_0 \eta^\alpha{}_\beta \phi^\beta
- a_1 \eta_{\beta(2)} \phi^{\alpha\beta(2)}. \nonumber 
\end{eqnarray}
{\bf $\zeta^{\alpha(2k+1)}$-transformations} For the fermions the
corrections look like:
\begin{eqnarray}
\delta \Phi^{\alpha(2k+1)} &=& \frac{b_k}{(2k+1)} 
\Omega^\alpha{}_\beta \zeta^{\alpha(2k)\beta} + \frac{a_k}{k(2k+1)}
\Omega^{\alpha(2)} \zeta^{\alpha(2k-1)} + a_{k+1} \Omega_{\beta(2)} 
\zeta^{\alpha(2k+1)\beta(2)}, \nonumber \\
\delta \Phi^\alpha &=& b_0 \Omega^\alpha{}_\beta \Phi^\beta +
a_1 \Omega_{\beta(2)} \zeta^{\alpha\beta(2)},
\end{eqnarray}
while for the spin-2 field we obtain:
\begin{eqnarray}
\delta H^{\alpha(2)} &=& \sum_{k=1}^{s-1} (-1)^{k+1} 
[ b_k \Phi^{\alpha\beta(2k)} \zeta^\alpha{}_{\beta(2k)}
+ a_k \Phi^{\alpha(2)\beta(2k-1)} \zeta_{\beta(2k-1)}
- a_k \Phi_{\beta(2k-1)} \zeta^{\alpha(2)\beta(2k-1)}] \nonumber \\
 && - b_0 \Phi^\alpha \zeta^\alpha + \frac{a_0}{8} 
( \phi^\alpha e^{\alpha\beta} \zeta_\beta + \phi_\beta 
e^{\beta\alpha} \zeta^\alpha). 
\end{eqnarray}
Note that in  this case the last formula is consistent with the
corrections to the spin 2 curvature:
\begin{eqnarray}
\Delta {\cal T}^{\alpha(2)} &=& \sum_{k=0}^{s-1} (-1)^{k+1}
\frac{b_k}{2} \Phi^{\alpha\beta(2k)} \Phi^\alpha{}_{\beta(2k)}
+ \sum_{k=1}^{s-1} (-1)^{k+1} a_k \Phi^{\alpha(2)\beta(2k-1)} 
\Phi_{\beta(2k-1)} \nonumber \\
 && + \frac{1}{16} (\phi_\beta e^{\beta\alpha} D \phi^\alpha -
\phi^\alpha e^\alpha{}_\beta D \phi^\beta) - \frac{a_0}{8}
(\Phi_\beta e^{\beta\alpha} \psi^\alpha - \Phi^\alpha
e^\alpha{}_\beta \phi^\beta) - \frac{3b_0}{4} E^{\alpha(2)}
\phi_\beta \phi^\beta. 
\end{eqnarray}

\subsection{Non-gravity 2}

In this case we have two  independent sets of the unfolded equations
with independent coupling constants $g_0$ and $\tilde{g}_0$. To see if
they can be consistent with the Lagrangian formulation, consider the
following ansatz for the cubic vertex:
\begin{eqnarray}
{\cal L}_1 &=& \sum_{k=0} (-1)^{k+1}[ \kappa_{1,k} 
\Phi_{\alpha(2k)\beta} \Omega^\beta{}_\gamma \Psi^{\alpha(2k)\gamma} -
\kappa_{2,k} \Phi_{\alpha(2k+1)\beta(2)} \Omega^{\beta(2)}
\Psi^{\alpha(2k+1)} \nonumber \\
 && \qquad \qquad + \kappa_{3,k} \Phi_{\alpha(2k+1)}
\Omega_{\beta(2)} \Psi^{\alpha(2k+1)\beta(2)}] \nonumber \\
 && + \rho_1 \Phi_\alpha (e^\alpha{}_\beta \Omega^\beta{}_\gamma +
\Omega^\alpha{}_\beta e^\beta{}_\gamma) \psi^\gamma + \rho_2
\Psi_\alpha (e^\alpha{}_\beta \Omega^\beta{}_\gamma +
\Omega^\alpha{}_\beta e^\beta{}_\gamma) \phi^\gamma \nonumber \\
 && + \rho_3 \phi_\alpha (e^\alpha{}_\beta \Omega^\beta{}_\gamma +
\Omega^\alpha{}_\beta e^\beta{}_\gamma) D \psi^\gamma + \rho_4
\psi_\alpha (e^\alpha{}_\beta \Omega^\beta{}_\gamma +
\Omega^\alpha{}_\beta e^\beta{}_\gamma) D \phi^\gamma \nonumber \\
 && + \phi_\alpha (\rho_5 E^\alpha{}_\beta \Omega^\beta{}_\gamma
+ \rho_6 \Omega^\alpha{}_\beta E^\beta{}_\gamma) \psi^\gamma. 
\end{eqnarray}
Consistency of such ansatz with the unfolded equations requires:
\begin{eqnarray}
\kappa_{1,k} &=& \frac{\tilde{a}_0}{a_0}(2k+1)h_{1,k}
= \frac{a_0}{\tilde{a}_0}(2k+1)\tilde{h}_{1,k}, \nonumber \\
\kappa_{2,k} &=& \frac{\tilde{a}_0}{a_0}(k+1)(2k+3)h_{2,k+1}
= \frac{a_0}{\tilde{a}_0} \tilde{h}_{3,k}, \\
\kappa_{3,k} &=& \frac{\tilde{a}_0}{a_0}h_{3,k} =
\frac{a_0}{\tilde{a}_0}(k+1)(2k+3)\tilde{h}_{2,k+1}. \nonumber
\end{eqnarray}
Besides, gauge invariance requires:
\begin{eqnarray}
\rho_1 &=& \frac{\tilde{a}_0M}{4}g_0, \qquad
\rho_2 = \frac{a_0}{4M}\tilde{g}_0, \nonumber \\
\rho_3 &=& \rho_4 = \frac{\rho_1}{2a_0} = \frac{\rho_2}{2\tilde{a}_0},
 \\
\rho_5 &=& \frac{a_0}{\tilde{a}_0}(2s+1)\tilde{g}_0, \qquad
\rho_6 = \frac{a_0}{\tilde{a}_0}(2s-1)\tilde{g}_0. \nonumber 
\end{eqnarray}
All these relations are satisfied provided
\begin{equation}
\tilde{a}_0{}^2M^2g_0 = a_0{}^2\tilde{g}_0.
\end{equation}
As in the previous case the Lagrangian is trivially invariant under
$\xi^{\alpha(2)}$-transformations.
{\bf $\eta^{\alpha(2)}$-transformations} Here we need the following
corrections:
\begin{eqnarray}
\delta \Phi^{\alpha(2k+1)} &=& - \frac{\tilde{a}_0}{a_0}
[h_{1,k} \eta^\alpha{}_\beta \Psi^{\alpha(2k)\beta} + h_{2,k}
\eta^{\alpha(2)} \Psi^{\alpha(2k-1)} + h_{3,k} \eta_{\beta(2)}
\zeta^{\alpha(2k+1)\beta(2)}, \nonumber \\
\delta \Phi^\alpha &=& - \frac{\tilde{a}_0}{a_0}
[h_{1,0} \eta^\alpha{}_\beta \Psi^\beta + h_{3,0} \eta_{\beta(2)}
\Psi^{\alpha\beta(2)}] + \rho_1 (e^\alpha{}_\beta
\eta^\beta{}_\gamma - \eta^\alpha{}_\beta e^\beta{}_\gamma)
\psi^\gamma, \\
\delta \phi^\alpha &=& - h_{1,0} \eta^\alpha{}_\beta \psi^\beta -
h_{3,0} \eta_{\beta(2)} \psi^{\alpha\beta(2)} \nonumber
\end{eqnarray}
and similarly for $\Psi$ with $\Psi \leftrightarrow \Phi$,
$a_0 \leftrightarrow \tilde{a}_0$ and $h \to \tilde{h}$. \\
{\bf $\tilde{\zeta}^{\alpha(2k+1)}$-transformations} The corrections
for the fermions have the form:
\begin{eqnarray}
\delta \Phi^{\alpha(2k+1)} &=& \frac{\tilde{a}_0}{a_0}
[h_{1,k} \Omega^\alpha{}_\beta \tilde{\zeta}^{\alpha(2k)\beta} +
h_{2,k} \Omega^{\alpha(2)} \tilde{\zeta}^{\alpha(2k-1)} + h_{3,k}
\Omega_{\beta(2)} \tilde{\zeta}^{\alpha(2k+1)\beta(2)}], \nonumber \\
\delta \Phi^\alpha &=& \frac{\tilde{a}_0}{a_0} [ h_{1,0}
\Omega^\alpha{}_\beta \tilde{\zeta}^\beta + h_{3,0} \Omega_{\beta(2)}
\tilde{\zeta}^{\alpha\beta(2)} ] 
\end{eqnarray}
and similarly for $\Psi$. While the $\zeta$-transformations for the
$H^{\alpha(2)}$ field can be extracted from the corrections to the
torsion:
\begin{eqnarray}
\Delta {\cal T}^{\alpha(2)} &=& \sum_{k=0}^{s-1}(-1)^{k+1}
[ \kappa_{1,k} \Phi^{\alpha\beta(2k)} \Psi^\alpha{}_{\beta(2k)}
+ \kappa_{2,k} \Phi^{\alpha(2)\beta(2k+1)} \Psi_{\beta(2k+1)}
\nonumber \\
 && \qquad \qquad + \kappa_{3,k} \Phi_{\beta(2k+1)}
\Psi^{\alpha(2)\beta(2k+1)} ] \nonumber \\
 && - \frac{\rho_1}{2} (\Phi_\beta e^{\beta\alpha} \psi^\alpha 
- \Phi^\alpha e^{\alpha\beta} \psi_\beta) 
 - \frac{\rho_2}{2} (\Psi_\beta e^{\beta\alpha} \phi^\alpha 
- \Psi^\alpha e^{\alpha\beta} \phi_\beta) \\
 && - \frac{\rho_3}{2} (\Phi_\beta e^{\beta\alpha} D\psi^\alpha 
- \phi^\alpha e^{\alpha\beta} D\psi_\beta) 
 - \frac{\rho_4}{2} (\psi_\beta e^{\beta\alpha} D\phi^\alpha 
- \psi^\alpha e^{\alpha\beta} D\phi_\beta) \nonumber \\
 && - \frac{\rho_5}{2} \phi_\beta E^{\beta\alpha} \psi^\alpha
 + \frac{\rho_6}{2} \phi^\alpha E^{\alpha\beta} \psi_\beta. \nonumber
\end{eqnarray}

\section*{Conclusion}

In this work we considered an interaction of the massless spin 2 field
with the massive higher spin fermions. The procedure begins with the
deformation of the unfolded equations in the sector of the gauge
invariant zero-forms where we have found three linearly independent
solutions. Then for all three cases we promoted this solutions to the
two other sectors and finally reconstructed corresponding cubic
vertices. Physically the most important case is of-course
gravitational interaction but we think that it is important that the
procedure works for all three cases. It may seems as not the shortest
way to the Lagrangian cubic vertices but such procedure allows us to
determine the minimum possible number of derivatives and in some sense
to resolve the ambiguity related with the field redefinitions. It
would be interesting to try to extend these approach to higher spins,
first of all to spins 5/2 and 3.

\end{document}